\documentclass{nature}

\usepackage{caption}
\usepackage{graphicx}
\usepackage[scaled]{helvet}
\usepackage{amsmath}
\usepackage{amssymb}
\usepackage{bm}
\usepackage{xfrac}
\newcommand*{\myfont}{\fontfamily{phv}\selectfont}
\usepackage[pdfstartpage=1]{hyperref}
\usepackage{anyfontsize}
\usepackage{ragged2e}

 \usepackage{color}

\definecolor{CLBlue}{rgb}{0, .25, .8}
\definecolor{MyBlue}{rgb}{0, .24, .40}
\definecolor{MyTurquoise}{rgb}{0, .53, .49}
\definecolor{MyGreen}{rgb}{0, .35, 0}
\definecolor{MyOrange}{rgb}{.8, .46, 0}
\definecolor{MyRed}{rgb}{.57, .07, 0}
\definecolor{MyPurple}{rgb}{.46, .1, .46}

\DeclareMathOperator{\Tr}{Tr}

\title{Quantifying the compressibility of the human brain}

\author{Nicholas J. Weaver$^{1,2,3}$, Joshua I. Faskowitz$^{4}$, Richard F. Betzel$^{5,6}$ \& Christopher W. Lynn$^{1,2,7,*}$}

\begin{document}

\maketitle

\begin{affiliations}
\item Department of Physics, Yale University, New Haven, CT 06520, USA
\item Quantitative Biology Institute, Yale University, New Haven, CT 06520, USA
\item Program in Physical and Engineering Biology, Yale University, New Haven, CT 06520, USA
\item Department of Psychological and Brain Sciences, Indiana University, Bloomington, IN 47405, USA
\item Department of Neuroscience, University of Minnesota, Minneapolis, MN 55455, USA
\item Masonic Institute for the Developing Brain, University of Minnesota, Minneapolis, MN 55414, USA
\item Wu Tsai Institute, Yale University, New Haven, CT 06520, USA
\item[*] Corresponding author: christopher.lynn@yale.edu
\end{affiliations}


\noindent {\large \myfont \textbf{Abstract}}
\vspace{-28pt}

\noindent\rule{\textwidth}{.5pt}

\noindent In the human brain, the allowed patterns of activity are constrained by the correlations between brain regions. Yet it remains unclear which correlations---and how many---are needed to predict large-scale neural activity. Here, we present an information-theoretic framework to identify the most important correlations, which provide the most accurate predictions of neural states. Applying our framework to cortical activity in humans, we discover that the vast majority of variance in activity is explained by a small number of correlations. This means that the brain is highly compressible: only a sparse network of correlations is needed to predict large-scale activity. We find that this compressibility is strikingly consistent across different individuals and cognitive tasks, and that, counterintuitively, the most important correlations are not necessarily the strongest. Together, these results suggest that nearly all correlations are not needed to predict neural activity, and we provide the tools to uncover the key correlations that are.

\newpage

\noindent {\large \myfont \textbf{Main}}
\vspace{-28pt}

\noindent\rule{\textwidth}{.5pt}

\noindent Non-invasive neuroimaging has revolutionized our ability to record large-scale activity in the human brain.\cite{Bassett2009HumanBrainNetworks} Individual regions, each reflecting the coarse-grained activity of a large population of neurons, interact to produce brain-wide activity that is responsible for learning, memory, perception, planning, and information processing.\cite{bressler2010large, park2013structural, rugg2013brain, bassett2011dynamic, fincham2002neural} The functional networks formed by these regions (nodes) and the correlations between them (edges) have provided key insights into human cognition.\cite{Lynn2019PhysicsOfBrainNetwork, Bassett2017NetworkNeuroscience, VANDENHEUVEL2010BrainNetwork} Variations in these networks are linked to cognitive disorders, human development, and environmental factors.\cite{Castellanos2013ClinicalApplications, dennis2014functional, baggio2014functional, uddin2010typical, chan2018socioeconomic, yang2016genetic} However, despite their crucial role in understanding cognition and disease, there remain basic questions about what correlations reveal about the nature of neural activity itself. Among all pairs of brain regions, which correlations are most important for predicting large-scale activity? And how many correlations do we need to construct an accurate model of the human brain?

Answering these questions is a problem of compression. The human brain is strongly correlated, with activity only exploring a small subset of all possible states.\cite{hipp2012large, jazayeri2021interpreting, mehrkanoon2014low} This suggests that a dense network of correlations is needed to constrain the neural activity. Alternatively, a small number of correlations may combine to have a large impact on the brain as a whole. In this simplified scenario, one would only need a sparse network of important correlations to predict the rest. If such a network exists, then the brain is highly compressible.

Information theory provides the foundation to make this intuition concrete.\cite{shannon1948mathematical, Cover&Thomas} Using the maximum entropy principle, one can map a network of correlations to predictions about neural activity.\cite{shannon1948mathematical, jaynes1957information} For a given number of correlations, the optimal network (which yields the most accurate predictions) provides the best compression of the system.\cite{Carcamo2025MinimaxReview} We show that this leads to a direct generalization of maximum entropy known as the minimax entropy principle,\cite{Zhu1997MinimaxTextureModeling, Lynn2025MinimaxPRR, Lynn2025MinimaxPRE, carcamo2024statisticalphysicslargescaleneural} which we solve to identify the optimal networks of correlations. Applying our framework to cortex-wide activity in healthy human subjects, we find that only a small number of correlations are needed to predict large-scale patterns of activity. We therefore discover that the brain is highly compressible. \\

\noindent {\myfont \large Constraining activity with a network of correlations}

\noindent Correlations tell us which regions are likely to activate together; once combined into a network they place complex constraints on the allowed patterns of activity. Consider $N$ brain regions with collective activity defined by the vector $\bm{x} = \{x_i\}$, where $x_i$ reflects the activity of region $i$. Any distribution over the states $P(\bm{x})$ defines a space of possible activity patterns. The size of this space---that is, our uncertainty about the neural states $\bm{x}$---is quantified by the entropy $S(P)$. In the simplest case, suppose we only measure the variance in activity $\sigma_i^2$ of each region in an experiment. The most unbiased distribution over states is composed  of independent Gaussians,\cite{jaynes1957information, dowson1973maximum} and our uncertainty about the neural activity is defined by the independent entropy $S_\text{ind} = \frac{1}{2}\sum_i \log \sigma_i^2 + \frac{N}{2}\log(2\pi e)$. At the other extreme, in addition to the individual variances, consider the full set of covariances between regions $\Sigma_{ij}$. By constraining the activity, these correlations reduce our uncertainty to the entropy of a fully-connected Guassian $S_\text{tot} = \frac{1}{2}\log |\Sigma| + \frac{N}{2}\log(2\pi e)$, where $|\Sigma|$ is the determinant of the covariance matrix.\cite{jaynes1957information, dowson1973maximum}

Between these two extremes, a subset of the covariances can be represented as a network $G$, with each edge $(ij)$ reflecting a covariance $\Sigma_{ij}$ included in our constraints.\cite{Lynn2019PhysicsOfBrainNetwork, Bassett2017NetworkNeuroscience} Given such a network, the most unbiased prediction for the distribution over states is the one with maximum entropy,\cite{jaynes1957information, dowson1973maximum}
\begin{equation}
\label{eq_PG}
P_G(\bm{x}) = \sqrt{\frac{|J|}{(2\pi)^N}} \exp \Big( -\frac{1}{2}\bm{x}^TJ\bm{x}\Big),
\end{equation}
where $J$ is the precision matrix. In practice, for each covariance in the network $G$, one must compute the precision $J_{ij}$ such that the model covariance $(J^{-1})_{ij}$ matches $\Sigma_{ij}$. This is equivalent to a Gaussian graphical model (GGM),\cite{Lauritzen1996GraphicalModels, Dyrba2019GGMAlzheimers, Zhang2019GGMSchizophrenia, Greenwald2017NIPS} for which there exist efficient inference algorithms (Methods).\cite{Uhler2017GaussianGraphicalModelsAlgebraic, Wermuth1977Algo} Thus, for a network of covariances $G$, our uncertainty about the collective activity is defined by the entropy $S_G = -\frac{1}{2}\log |J| + \frac{N}{2}\log(2\pi e)$.

\begin{figure}[t!]
\centering
\includegraphics[width = .95\textwidth]{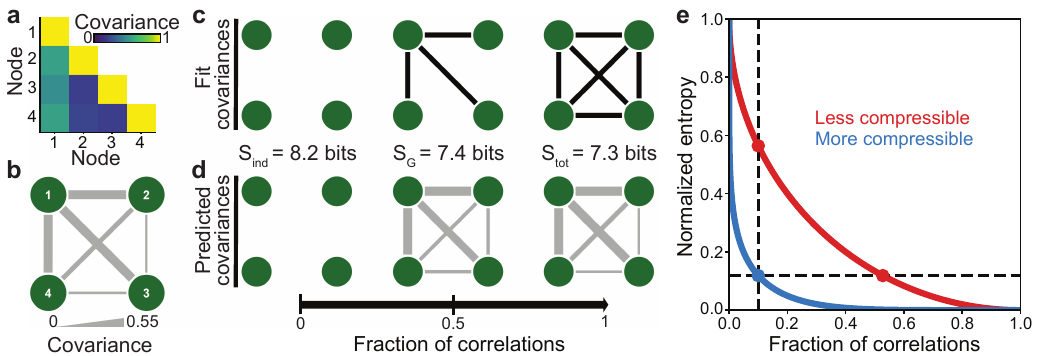} \\
\raggedright
\captionsetup{labelformat=empty}
{\spacing{1.25} \caption{\small \textbf{Fig.~\ref{fig:intro} $|$ Quantifying uncertainty given a network of correlations.} \textbf{a-b}, Covariance matrix (\textbf{a}) and network representation (\textbf{b}) for a minimal system of four regions with z-scored activity (Methods). \textbf{c-d}, Networks of correlations $G$ (\textbf{c}) and the covariances predicted by the corresponding maximum entropy models $P_G$ (\textbf{d}). The network formed by the three strongest covariances produces accurate predictions for all covariances (middle). \textbf{e}, Illustration of the normalized entropy $\tilde{S}$ versus the fraction of correlations (or the density of the network $G$) for neural activity that is either more compressible (blue) or less compressible (red). In a more compressible system, one can achieve lower uncertainty for a given number of correlations (vertical line), and one can find a sparser network (with fewer correlations) to achieve a given level of uncertainty (horizontal line). \label{fig:intro}}}
\end{figure}

Consider a minimal example of four brain regions (Fig.~\ref{fig:intro}a-b). With no covariances, the regions behave independently with high uncertainty $S_\text{ind}$ (Fig.~\ref{fig:intro}c, \textit{left}). With all of the covariances, we necessarily capture all of the correlation structure, reducing our uncertainty to $S_\text{tot}$ (Fig.~\ref{fig:intro}c, \textit{right}). Between these limits, one might hope to find a small subset of covariances that produces a large reduction in uncertainty; that is, a good compression. Indeed, by constraining the three strongest covariances, the maximum entropy model accurately predicts the remaining indirect correlations (Fig.~\ref{fig:intro}d), and the entropy $S_G$ nearly reduces to that of the fully-connected network $S_\text{tot}$.

We therefore arrive at the following picture: As one includes more correlations in a network, the uncertainty about neural activity decreases, resulting in a hierarchy of entropy $S_\text{ind} \ge S_G \ge S_\text{tot}$. To compare across different datasets, we define the normalized entropy,
\begin{equation}
\tilde{S}_G = \frac{S_G - S_\text{tot}}{S_\text{ind} - S_\text{tot}},
\end{equation}
which quantifies the progress from no correlations ($\tilde{S}_G = 1$) to a network that explains all of the structure in the neural activity ($\tilde{S}_G = 0$). If the brain is easier to compress, then for a given number of correlations, one should be able to find a network that reduces our uncertainty $\tilde{S}_G$ (Fig.~\ref{fig:intro}e, \textit{vertical line}). Equivalently, to achieve a given level of uncertainty, one should be able to find a network with fewer correlations (Fig.~\ref{fig:intro}e, \textit{horizontal line}). Thus, in order to quantify the compressibility of the human brain, we must first identify the optimal networks of correlations. \\

\noindent {\myfont \large Optimally compressing neural activity}

\noindent For a given number of correlations, we seek the network $G$ that minimizes the entropy $S_G$. This is an instance of the ``minimax entropy" principle, which has recently been proposed to construct optimal models of complex systems, but has yet to be applied to human neural activity.\cite{Zhu1997MinimaxTextureModeling, Carcamo2025MinimaxReview, Lynn2025MinimaxPRE, Lynn2025MinimaxPRR, carcamo2024statisticalphysicslargescaleneural} Notably, since $P_G$ is a maximum entropy model [Eq.~(\ref{eq_PG})], we show that minimizing the entropy $S_G$ is equivalent to minimizing the KL divergence between the model and the data (Methods). Thus, the optimal network $G$, which minimizes our uncertainty, also provides the most accurate description of the observed activity. 

Given a desired number of correlations, identifying the optimal network poses two distinct challenges. First, for a given network $G$, one must compute the model $P_G$ that matches the covariances $\Sigma_{ij}$ in $G$; this can be accomplished using efficient GGM algorithms (Methods).\cite{Uhler2017GaussianGraphicalModelsAlgebraic, Wermuth1977Algo} Second, one must search over all networks $G$ to find the one that provides the best compression, minimizing the entropy $S_G$. However, the number of possible networks explodes combinatorially, making brute force search impossible. Instead, as is common in discrete optimization, we decompose the search into a sequence of local optimization problems.\cite{Cormen2009IntroAlgos, Carcamo2025MinimaxReview} Beginning with the empty network, we iteratively add the correlation that produces the largest decrease in entropy $S_G$, thus maximally reducing our uncertainty. When these drops in entropy become small, we can even expand in the weak-correlation limit analytically to further increase efficiency (Methods). We repeat this greedy algorithm until all correlations have been added. In doing so, we identify not only a single (locally) optimal network, but the sequence of all optimal compressions across all numbers of correlations. The result is a compression curve $\tilde{S}(f)$, which defines the minimum uncertainty that can be achieved with a given fraction of the correlations $f$ (Fig.~\ref{fig:intro}e).

\begin{figure}
\centering
\includegraphics[width = 0.8\textwidth]{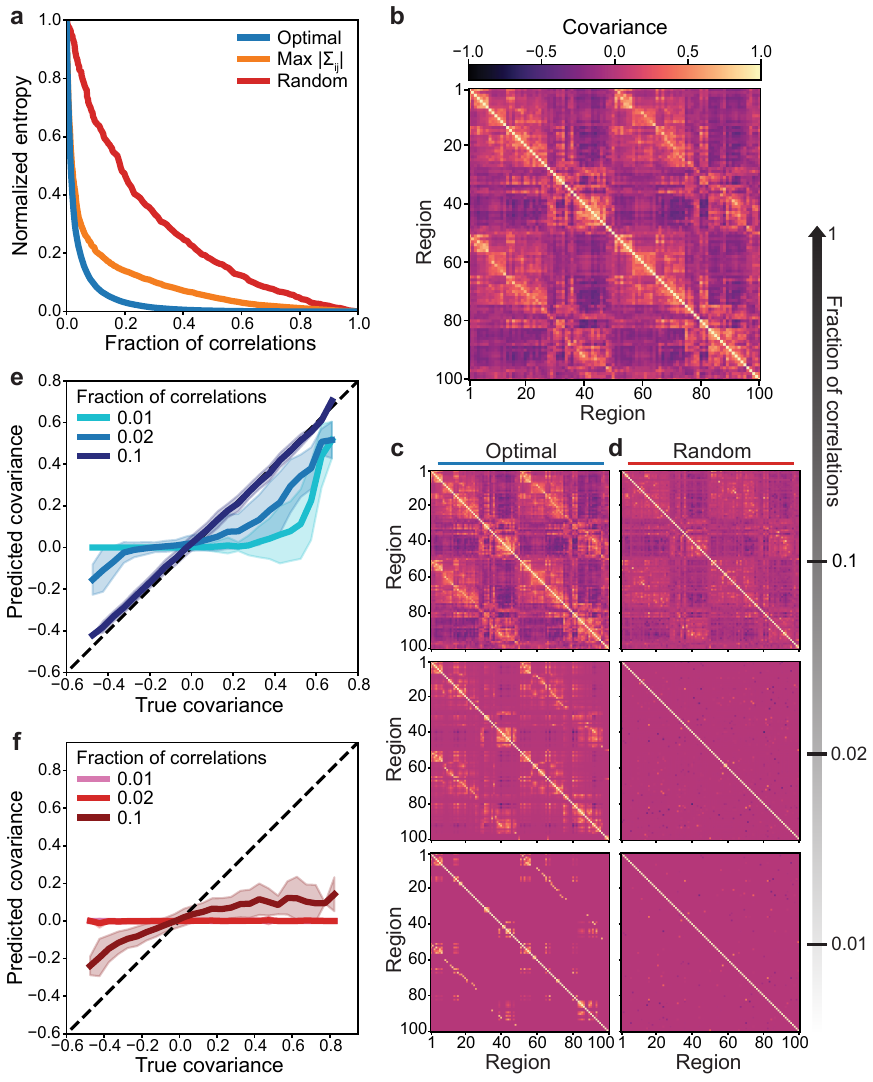} \\
\raggedright
\captionsetup{labelformat=empty}
{\spacing{1.25} \caption{\small \textbf{Fig.~\ref{fig:results1} $|$ Human neural activity is highly compressible.} \textbf{a}, Normalized entropy $\tilde{S}_G$ as a function of the fraction of correlations in $G$ for the optimal networks (blue), strongest correlations $|\Sigma_{ij}|$ (orange), and random networks (red). \textbf{b}, Covariances between brain regions measured in activity concatenated across all subjects and tasks. \textbf{c-d}, Predicted covariances based on the optimal networks (\textbf{c}) and random networks (\textbf{d}) with different fractions of the correlations. \textbf{e-f}, Predicted covariances versus their true values for models constructed from optimal networks (\textbf{e}) and random networks (\textbf{f}). Lines and shaded regions indicate means and standard deviations across all covariances that are not included in each model. \label{fig:results1}}}
\end{figure}

We apply our framework to cortex-wide activity from 99 healthy adults both at rest and across a suite of seven cognitive tasks, recorded using functional magnetic resonance imaging (fMRI) as part of the Human Connectome Project.\cite{Van2013HumanConnectomeProject} The collective activity consists of blood-oxygen-level–dependent (BOLD) fMRI signals from $N = 100$ cortical parcels.\cite{SchaeferYeo2018} Concatenating this activity across all subjects and tasks, we apply our minimax entropy algorithm to compute optimal compressions across all numbers of correlations. Strikingly, we find that with only a small number of correlations, the entropy drops significantly; only 1.4\% of the correlations are needed to achieve a 50\% reduction in entropy, and a 90\% reduction only requires 9\% of the correlations (Fig.~\ref{fig:results1}a). Intuitively, one might expect the strongest correlations between regions to provide the tightest constraints on neural activity. To test this hypothesis, we compare against networks consisting of the strongest covariances. While these networks provide significantly better compressions than random correlations, the optimal networks still achieve up to orders of magnitude lower uncertainty (Fig.~\ref{fig:results1}a). Together, these results indicate that the human brain is highly compressible, and that the optimal compressions themselves do not simply consist of the strongest correlations. 

Given this compressibility, with only a small number of important correlations, we should be able to predict the entire correlation structure (Fig.~\ref{fig:results1}b). For the optimal networks, as we increase the number of correlations, we see that the full structure quickly comes into focus (Fig.~\ref{fig:results1}c). In fact, by fitting only 10\% of the correlations between regions, the optimal model $P_G$ quantitatively predicts the remaining 90\% (Fig.~\ref{fig:results1}e). By contrast, with the same number of random correlations, most of the structure is lost (Fig.~\ref{fig:results1}d), and the model significantly underestimates the true strengths of correlations (Fig.~\ref{fig:results1}f). Thus, while the brain is strongly correlated (Fig.~\ref{fig:results1}b), only a sparse network of these correlations is needed to explain the rest. \\

\noindent {\myfont \large Quantifying the compressibility of neural activity}

\begin{figure}[t!]
\centering
\includegraphics[width = 0.75\textwidth]{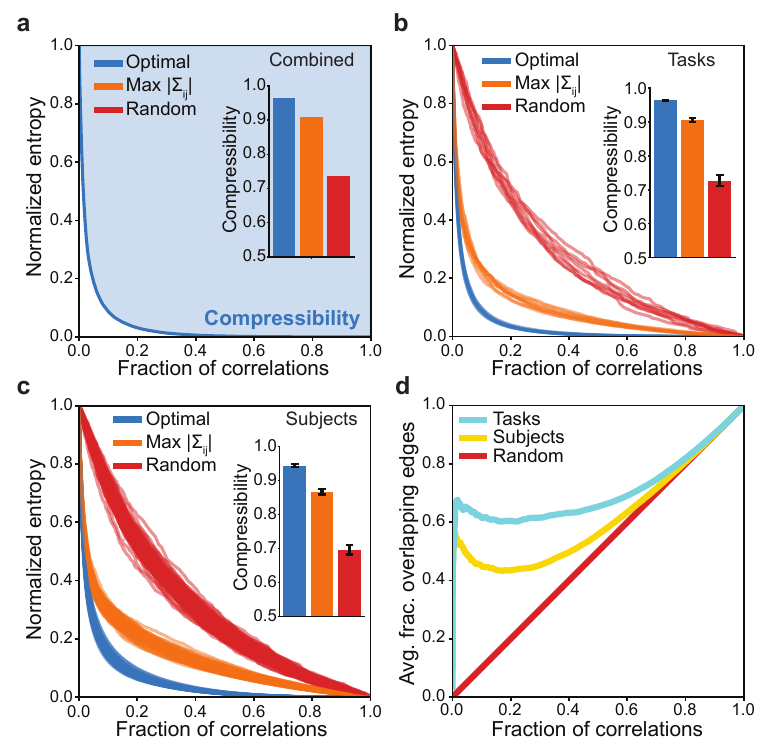} \\
\raggedright
\captionsetup{labelformat=empty}
{\spacing{1.25} \caption{\small \textbf{Fig.~\ref{fig:results3} $|$ Quantifying compressibility across subjects and cognitive tasks.} \textbf{a}, Optimal normalized entropy $\tilde{S}(f)$ versus the fraction of correlations $f$ for data combined across all tasks and subjects. Compressibility is the shaded area above the compression curve. Inset displays compressibility values for optimal, maximum correlation, and random networks. \textbf{b-c}, Normalized entropy versus fraction of correlations in optimal (blue), maximum correlation (orange), and random (red) networks for data within specific tasks (\textbf{b}) and within specific subjects (\textbf{c}). Insets display compressibility values averaged across tasks (\textbf{b}) and across subjects (\textbf{c}) with one-standard-deviation error bars. \textbf{d}, Average overlap (fraction of shared correlations) between optimal networks for pairs of tasks (blue) and pairs of subjects (yellow) versus the fraction of correlations. Red line illustrates the overlap between random networks. \label{fig:results3}}}
\end{figure}

\noindent We are now prepared to quantify the compressibility of the brain. Rather than choosing a specific number of correlations, we would like compressibility to be a property of the neural activity itself. We therefore define compressibility to be the amount of uncertainty that can be removed via compression, averaged across all numbers of correlations,
\begin{equation}
C = 1 - \int_0^1 \tilde{S}(f)df.
\end{equation}
Visually, the compressibility represents the area above the compression curve (Fig.~\ref{fig:results3}a). This approaches $C=1$ for perfectly compressible activity (with all structure concentrated on one correlation) and $C = 0$ for perfectly incompressible activity (with all correlations needed to explain the structure). For the fMRI data combined across subjects and tasks, we find a compressibility $C = 0.96$ near the theoretical maximum (Fig.~\ref{fig:results3}a); meanwhile, sub-optimal networks (composed of random or strong correlations) yield lower compressibilities. This means that, on average across all numbers of correlations, one can achieve a 96\% reduction in uncertainty about the neural activity.

One might be concerned that concatenating activity across subjects and tasks introduces spurious correlations that lead to good compressions. Instead, we can investigate the compressibility of the brain within a given cognitive task (combined across subjects) or for a specific subject (combined across tasks, including rest). In both cases, our compression framework still discovers sparse networks of correlations that produce steep drops in entropy (Fig.~\ref{fig:results3}b-c). Averaging across all numbers of correlations, we find that the brain is highly compressible both within tasks ($C = 0.96$) and for individual subjects ($C = 0.94$). Moreover, we see that the compression curves are surprisingly consistent across different tasks (Fig.~\ref{fig:results3}b) and subjects (Fig.~\ref{fig:results3}c), suggesting that the optimal networks themselves may be consistent. Indeed, for the top 10\% of correlations, we find that the optimal networks for specific subjects overlap 53 times more strongly than random, and this overlap increases to 62 times more than random for different cognitive tasks (Fig.~\ref{fig:results3}d). Given this consistency, and given that they capture the vast majority of correlations, what is the network structure of optimal compressions? \\ 

\noindent {\myfont \large Structure of optimal correlations}

\begin{figure}[t!]
\centering
\includegraphics[width = .85\textwidth]{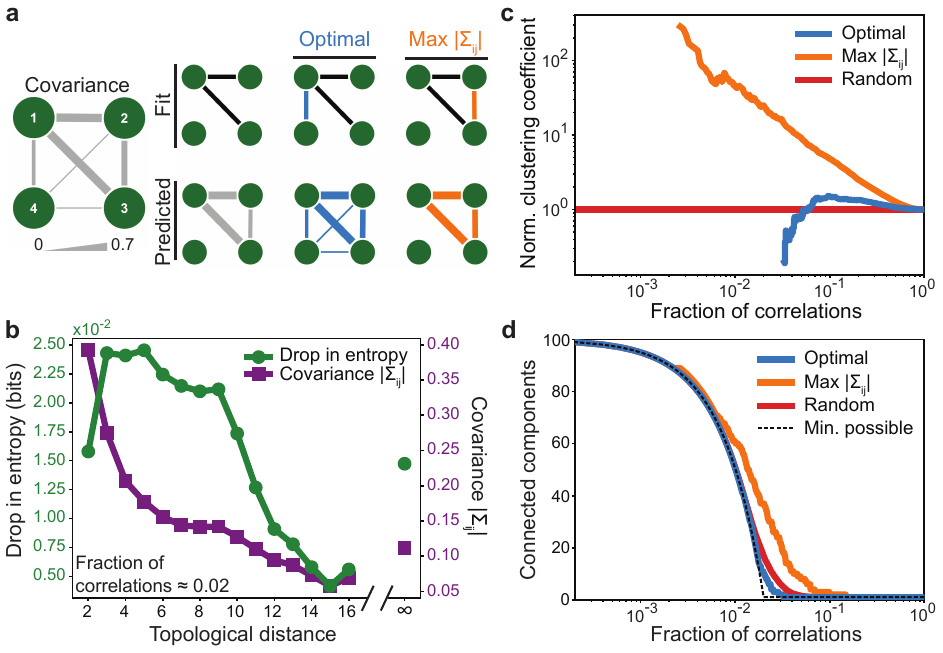} \\
\raggedright
\captionsetup{labelformat=empty}
{\spacing{1.25} \caption{\small \textbf{Fig.~\ref{fig:results2} $|$ Network structure of optimal correlations.} 
\textbf{a}, For a minimal system of four regions, the first two optimal covariances are the strongest, yielding an accurate prediction for the third strongest (\textit{left}). Thus, the third strongest covariance is redundant, providing little improvement in accuracy (\textit{right}). The third optimal covariance is weaker, but provides more accurate predictions (\textit{middle}). \textbf{b}, For the optimal model with 100 correlations, we plot the average drop in entropy from fitting covariances and average strength of covariances as a function of the topological distance between brain regions in the network (that is, the length of the shortest path). Topological distance is infinity if there does not exist a path between regions. \textbf{c}, Global clustering coefficient (normalized by the density of the network) versus the fraction of correlations. Note that zero values are not shown. \textbf{d}, Number of connected components versus the fraction of correlations, where each random value is averaged over 100 random networks. Dashed line indicates the minimum possible value. In \textbf{a-c}, neural data is combined across all tasks and subjects. \label{fig:results2}}}

\end{figure}

\noindent We have seen that the optimal correlations between regions, which minimize our uncertainty about neural activity, are not simply the strongest (Figs.~\ref{fig:results1}a and \ref{fig:results3}b-c). To understand how this is possible, consider the minimal system in Fig.~\ref{fig:results2}a. With knowledge of the two strongest covariances $\Sigma_{12}$ and $\Sigma_{13}$, one can accurately predict the next strongest covariance $\Sigma_{23}$. Thus, when selecting the third correlation to include in the model, rather than choosing the strongest option $\Sigma_{23}$, one can gain more information by selecting a weaker covariance (in this case $\Sigma_{14}$). In general, while strong correlations tend to form tight loops,\cite{masuda2018clustering} some of these may be predicted indirectly from the others; these strong but redundant correlations do not reduce our uncertainty about the neural activity.

This phenomenon plays a key role in shaping the network structure of optimal compressions. For example, for the optimal network with 100 correlations, we find that the strength of correlations decreases with increasing distance in the network (Fig.~\ref{fig:results2}b). This means that selecting the strongest correlations leads to networks with many short loops, a property known as clustering.\cite{Lynn2019PhysicsOfBrainNetwork} Meanwhile, the largest drops in entropy are not achieved by these strong, short-range correlations, but instead by correlations of intermediate strength and distance in the network (Fig.~\ref{fig:results2}b). As a result, while the strongest correlations form networks with high clustering, the optimal networks exhibit orders of magnitude lower clustering (Fig.~\ref{fig:results2}c). Rather than concentrating connectivity within tight clusters of regions, we find that optimal networks spread connectivity across distant parts of the network. This leads to the formation of a single giant component---wherein each region connects at least indirectly to every other region---with nearly as few correlations as possible (Fig.~\ref{fig:results2}d).\cite{newman2018networks, molloy1995critical}

\begin{figure}[t!]
\centering
\includegraphics[width = .7\textwidth]{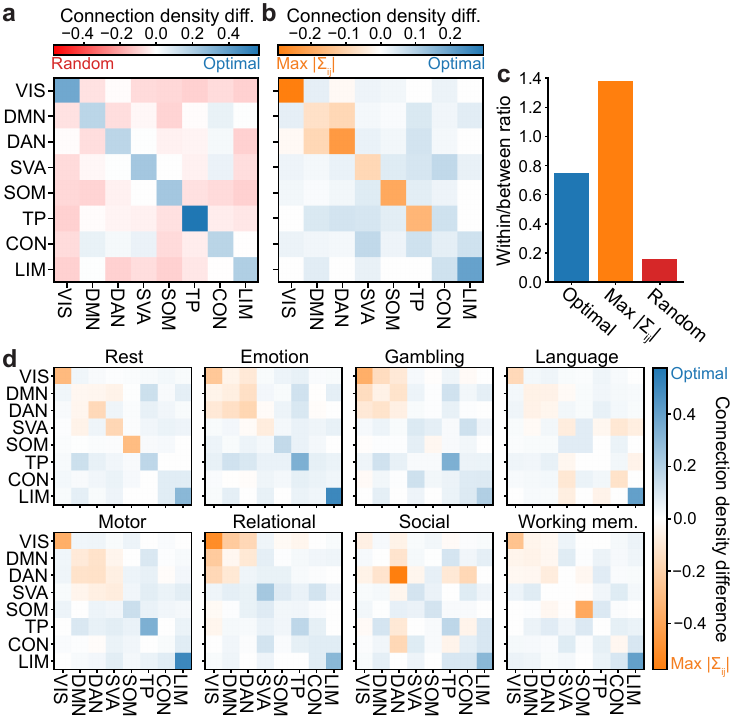} \\
\raggedright
\captionsetup{labelformat=empty}
{\spacing{1.25} \caption{\small \textbf{Fig.~\ref{fig:results4} $|$ Systems-level structure of optimal compressions.} \textbf{a}, Difference in the density of connections between the optimal network (blue) and random networks (red) for pairs of regions in different cognitive systems. Diagonal (off-diagonal) elements indicate connections within (between) systems. \textbf{b}, Difference in the density of connections between the optimal network (blue) and the network of strongest correlations (orange). \textbf{c}, Ratio of correlations within versus between cognitive systems. In \textbf{a-c}, neural data is combined across all tasks and subjects. \textbf{d}, Difference in the density of connections between optimal and maximum correlation networks during specific cognitive tasks including rest (data combined across subjects). Across all panels, networks contain 10\% of correlations. We coarse-grain regions into eight systems: visual (VIS), default mode (DMN), dorsal attention (DAN), salience/ventral attention (SVA), somatomotor (SOM), temporoparietal (TP), control (CON), and limbic (LIM).
\label{fig:results4}}}
\end{figure}

Highly correlated brain regions are known to form functionally specific subnetworks known as cognitive systems.\cite{Lynn2019PhysicsOfBrainNetwork, Bassett2017NetworkNeuroscience, crossley2013cognitive} Our above results suggest that, rather than focusing on strong correlations within these systems, one might construct better models of neural activity by including weaker correlations between different cognitive systems. To test this hypothesis, we sort the 100 regions into eight subnetworks \cite{Yeo2011,SchaeferYeo2018} Correlations within each of these cognitive systems are much more likely to appear in optimal networks than random, indicating that they are important for constraining neural activity (Fig.~\ref{fig:results4}a). However, relative to the strongest correlations, optimal networks tend to connect regions spanning different cognitive systems, suggesting that many of the strong correlations within systems are redundant (Fig.~\ref{fig:results4}b). In fact, while the majority of the strongest correlations lie within specific systems, most of the optimal correlations span separate subnetworks (Fig.~\ref{fig:results4}c).

We observe subtle deviations from this pattern in different cognitive tasks (Fig.~\ref{fig:results4}d): Correlations between the visual (VIS), default mode (DMN), and dorsal attention (DAT) systems, which are highly engaged during visual and motor response tasks,\cite{raichle2001default, szczepanski2013functional, desimone1995neural} provide less information about cortex-wide activity than one would expect from their strength alone. Meanwhile, correlations within the limbic (LIM) and temporoparietal (TEP) systems, which are associated with emotional and social processing,\cite{catani2013revised, krall2015role} are weak but important for constraining activity. Yet across all tasks, we consistently find that optimal compressions tend to spread connectivity between different cognitive systems (Fig.~\ref{fig:results4}d). Together, these results demonstrate how, by avoiding tight clusters of strong but redundant correlations, one can achieve a much more effective compression of the human brain. \\

\noindent {\myfont \large Discussion}

\noindent In the human brain, the correlations between brain regions have provided foundational insights into cognition and disease.\cite{Bassett2009HumanBrainNetworks, bressler2010large, park2013structural, rugg2013brain, bassett2011dynamic, fincham2002neural, Lynn2019PhysicsOfBrainNetwork, Bassett2017NetworkNeuroscience, VANDENHEUVEL2010BrainNetwork, Castellanos2013ClinicalApplications, dennis2014functional, baggio2014functional, uddin2010typical, chan2018socioeconomic, yang2016genetic} But how many correlations---and which correlations---are needed to explain brain-wide activity remains poorly understood. To identify the most important correlations between brain regions, which generate the most accurate predictions of neural states, we show that one must solve the minimax entropy problem, which has roots in information theory and statistical physics.\cite{Carcamo2025MinimaxReview, Zhu1997MinimaxTextureModeling, Lynn2025MinimaxPRR, Lynn2025MinimaxPRE, carcamo2024statisticalphysicslargescaleneural} In fMRI data from a large cohort of subjects across several cognitive tasks,\cite{Van2013HumanConnectomeProject} we discover that only a small number of correlations are needed to explain the observed cortex-wide activity (Fig.~\ref{fig:results1}). This reveals that the brain is highly compressible: one can ignore nearly all of the correlations between regions while still accurately predicting neural activity. In fact, we demonstrate that the compressibility of the human brain is near the theoretical maximum, and that this high compressibility is consistent across different subjects and cognitive tasks (Fig.~\ref{fig:results3}). Finally, we find that the most important correlations for predicting neural activity are not simply the strongest, a fact that has profound implications for the structure of optimal compressions (Figs.~\ref{fig:results2} and \ref{fig:results4}).

These results demonstrate that the constraints on neural activity in the human brain are distributed highly non-uniformly. A small fraction of the correlations between regions have orders of magnitude more influence on brain-wide activity than the vast majority of correlations. In turn, this suggests that the human brain may be dominated by an extremely sparse backbone of abnormally strong interactions. Indeed, such heavy-tailed distributions of connection strength are observed in the structural wiring between brain regions.\cite{roberts2016contribution, gast2024weighting, cirunay2025scale} Identifying the interactions that dominate neural activity---which we have shown differ systematically from the strongest correlations---may have important implications for understanding neurological disorders and developing targeted treatments.\cite{Lynn2019PhysicsOfBrainNetwork, breit2004deep, kobayashi2003transcranial, betzel2016optimally}

Moreover, with the tools to quantify the compressibility of the human brain, a number of questions immediately arise. How does the brain's high compressibility emerge during development\cite{luna2010has, dosenbach2010prediction}? And how is it altered in diseases such as Alzheimer's,\cite{wang2007altered} Parkinson's,\cite{wolters2019resting} and Schizophrenia\cite{bassett2012altered}? Moreover, how does the network and systems-level structure of optimal compressions vary across development and disease? With the explosion of available fMRI data, one can directly apply our compression framework to begin answering these questions. In fact, our methods can be used to study any continuous-valued recordings of neural activity, opening the door for future investigations of neural activity across different imaging modalities (including EEG and MEG\cite{chang2013eeg, brookes2011measuring}) and even distinct species.\cite{stephan2000computational, white2011imaging} Given the tendency of direct interactions to produce indirect correlations (Fig.~\ref{fig:results2}a), we anticipate that many strong correlations in the brain may in fact be redundant, the natural consequences of neighboring interactions. In turn, we hypothesize that many neural systems will be amenable to simplified descriptions with only a small number of important correlations; that is, highly compressible.

\newpage

\noindent {\large \myfont \textbf{Methods}}
\vspace{-28pt}

\noindent\rule{\textwidth}{.5pt}

\begin{methods}
\setlength{\parindent}{0pt}

\subsection{Maximum entropy.}
Consider the matrix of covariances $\Sigma$ between $N$ brain regions. A subset of these covariances can be represented as a network $G$ (Fig.~\ref{fig:intro}). Given a network of correlations, the most unbiased prediction for the remaining covariances is provided by the maximum entropy model in Eq.~(\ref{eq_PG}). To compute the maximum entropy model, for each edge $(ij)$ in the network $G$, one must compute the entry of the precision matrix $J_{ij}$ such that the corresponding model covariance $(J^{-1})_{ij}$ matches the experimental value $\Sigma_{ij}$. There exist efficient algorithms for solving this problem, which are guaranteed to converge to the correct solution\cite{Uhler2017GaussianGraphicalModelsAlgebraic,Wermuth1977Algo,Speed1986ConvProofs} Here, we use Algorithms 1 and 2 in Ref. \citeonline{Uhler2017GaussianGraphicalModelsAlgebraic}. Algorithm 1 iterates over entries of the precision matrix corresponding to edges in the network $G$, while Algorithm 2 iterates over elements of the model covariance corresponding to edges not in $G$. Thus, Algorithm 1 (Algorithm 2) is more efficient for networks with fewer (more) than half the available edges. When calculating maximum entropy models, we switch from Algorithm 1 to 2 at 1500 correlations, and we check that solutions converge within experimental errors (Supplementary Information).

\subsection{Minimax entropy.}

Given a specified number of correlations, there are many different subsets---or networks---one could choose. Here we show that the optimal network, which provides the most accurate model of the neural activity, also generates the best compression of the data. Letting $Q(\bm{x})$ denote the experimental distribution over states $\bm{x}$, the Kullback-Leibler (KL) divergence\cite{Cover&Thomas} with a maximum entropy model $P_G(\bm{x})$ simplifies to a difference in entropy,\cite{Carcamo2025MinimaxReview}
\begin{align}
    D_{\text{KL}}(Q||P_G) &= \left\langle \log\frac{Q}{P_G}\right\rangle_{Q} \\
    &= \langle \log Q\rangle_{Q}  - \langle \log P_G\rangle_{Q} \\
    &= -S(Q) + \frac{N}{2}\log (2\pi) - \frac{1}{2}\log(|J|) + \sum_{(ij)\in G} J_{ij} \langle x_i x_j \rangle_{Q} \\
    &= -S(Q) + \frac{N}{2}\log (2\pi) - \frac{1}{2}\log(|J|) +\sum_{(ij)\in G} J_{ij} \langle x_i x_j \rangle_{P_G} \\
    &= -S(Q) - \langle \log P_G \rangle_{P_G} \\
    &= S_G - S(Q),
\end{align}
where $S(Q)$ is the data entropy. We therefore find that minimizing the KL divergence is equivalent to minimizing the entropy $S_G$ of the maximum entropy model $P_G$. This is the minimax entropy principle tells us that the optimal network $G$, which provides the most accurate description of the activity, also produces the best compression of the data.

For a given number of correlations, the number of possible networks explodes combinatorially, making a brute force search for the optimal network impossible. Mathematically, this optimization is equivalent to constructing a Gaussian graphical model (GGM) with an $L_0$ regularization on the precision matrix, which is notoriously difficult. One heuristic for overcoming this challenge is using the graphical lasso, with an $L_1$ norm on the inverse covariance to penalize a lack of sparsity.\cite{Friedman2007GLASSO, Varoquaux2013LearningFunctionalConnectomes, NIPS2010BrainCovarianceSelection, Banerjee2006L1Lasso} Here, we instead decompose the problem into a sequence of local optimizations that we can solve exactly. Specifically, we develop a greedy algorithm for constructing the optimal network $G$. Starting from the independent model (with no correlations), we iteratively add the correlation to our model that produces the largest drop in entropy $S_G$. Using this greedy approach, we can identify the locally optimal network not only for a one number of correlations, but across all numbers of correlations, thus generating the entire entropy curve (Fig.~\ref{fig:results1}). We confirm that this greedy algorithm outperforms many other heuristics (Supplementary Information).

\subsection{Improving the efficiency of compression.}

At each step of the greedy algorithm, we must fit $O(N^2)$ different maximum entropy models: one for each of the possible correlations that could be added to the model. This process must then be repeated for each of the $O(N^2)$ steps of the greedy algorithm. To scale our compression to large-scale data, we therefore need strategies for improving efficiency. To begin, we note that if the network has no loops, then each correlation produces a drop in entropy equal to the mutual information between the two brain regions.\cite{Nguyen2017EntandMI,Lynn2025MinimaxPRE,Lynn2025MinimaxPRR} Thus, for sparse networks with no loops, one can simply select the correlations corresponding to the largest mutual information.

Once loops are included in the network, we require new strategies for efficiency. For this purpose, we derive an analytic approximation to the drop in entropy that does not require fitting a new maximum entropy model (Supplementary Information). We use perturbation theory to estimate entropy drops in the limit of small errors on the predicted correlations (or, equivalently, in the limit of small precision entries). As the greedy algorithm progresses, and the network becomes denser, the entropy drops become very small (Fig.~\ref{fig:results1}a), making our approximation accurate. In practice, we compute entropy drops exactly for the first 50 correlations before switching to our approximation. We confirm the accuracy of our approximation, even in sparse networks with only 2\% of correlations (Supplementary Information).

\subsection{Data.} 
We use fMRI data from the Human Connectome Project (HCP) from Ref. \citeonline{Van2013HumanConnectomeProject} for 99 human subjects. The HCP study was approved by the Washington University Institutional Review Board and informed consent was obtained from all subjects. Participants were not compensated. The subjects were scanned at rest and while performing seven cognitive tasks: emotion, gambling, language, motor, relational, social, and working memory. When we refer to ``tasks'', we include rest for convenience. A comprehensive description of the imaging parameters and image preprocessing can be found in Ref.~\citeonline{Glasser2013fMRI}. Using a 3T Siemens Connectome Skyra with a 32-channel head coil, gradient-echo EPI images were acquired during eight conditions with the following parameters: TR = 720 ms; TE = 33.1 ms; 2-mm isotropic voxel resolution; flip angle = 52°; multiband acceleration factor = 8. See Ref.~\citeonline{Barch2013fMRI} for details about the details and timing of each administered task. Minimally preprocessed data, which included slice timing, motion, and distortion correction, normalized to a common surface space template were downloaded. In addition, time series were linearly detrended, temporally filtered (0.008-0.08 Hz), and had 24 head motion parameters, eight mean signals from white matter and cerebrospinal fluid, and four global signals regressed out of the time series data, corresponding to strategy six in Ref.~\citeonline{Parkes2018fMRI}. Denoised vertex-wise time series were averaged within each area of the Schaefer 100 parcellation\cite{SchaeferYeo2018} at each time step, to create parcel-wise time series.

We concatenate scans for each combination of subject and task and z-score them, subtracting the mean and dividing by the standard deviation. Data was combined into larger timeseries in three ways: across all tasks and subjects (yielding combined data), across tasks for a given subject (yielding subject data), and across subjects for a given task (yielding task data). Note that all activity was mapped to the same atlas, making it meaningful to combine the time series from the same region in different subjects.\cite{Yeo2011,SchaeferYeo2018} Scans for different tasks have different lengths, with the shortest (emotion) having just 176 frames. There were two scans per task per subject, and a maximum of four for rest. For task and subject data, two subjects were excluded due to having too few rest scans. These two subjects still had their data included in the combined data. For task data, to avoid differences due to sampling, we only use the beginning of each scan with the same length as that of the shortest task. For each task, two scans (including rest) were included per subject. Subject data included two scans per non-rest task and three rest scans per subject. This ensured the same amount of data for all subjects.

\end{methods}

\section*{Data Availability}

The data analyzed in this paper are openly available at: \\https://github.com/NicholasJWeaver/BrainCompressibility2025/

\section*{Code Availability}

The code used to perform the analyses in this paper is openly available at:\\
https://github.com/NicholasJWeaver/BrainCompressibility2025/


\newpage
\begin{addendum}

\item[Supplementary Information.] Supplementary text and figures accompany this paper.

\item[Acknowledgements.] We thank Carlton Smith, David Carcamo, Jose Betancourt, Matt Leighton, Benjamin Machta, and Damon Clark for enlightening discussions and comments on earlier versions of the paper. N.J.W. and C.W.L. acknowledge support from the National Institutes of Health (NIH/NIGMS R35GM160188), as well as the Department of Physics, the Quantitative Biology Institute, and the Physical and Engineering Biology Program at Yale University. R.F.B. acknowledges support from the National Science Foundation (2023985), National Institute of Aging (AG075044), the National Institute on Drug Abuse (NS125026), and MNDrive Brain Conditions.
 
\item[Author Contributions.] N.J.W. and C.W.L. conceived the project, designed the models, and wrote the paper. N.J.W. performed the analysis. J.I.F. and R.F.B. processed the data and provided input on analyses and writing.
 
\item[Competing Interests.] The authors declare no competing financial interests.
 
\item[Corresponding Author.] Correspondence and requests for materials should be addressed to C.W.L. \\(christopher.lynn@yale.edu).
 
\end{addendum}







\setcounter{figure}{0}

\noindent {\large \myfont \textbf{Supplementary Materials}}



\section{Speeding up the minimax entropy procedure, with additional details.} 
We need to find ways to speed up the greedy algorithm for building minimax entropy models. We can first use the result that if adding an edge $(ij)$ does not create any loops in the graph, the entropy drop from adding that edge is simply the mutual information between the nodes in that edge:~\cite{Nguyen2017EntandMI,Lynn2025MinimaxPRE,Lynn2025MinimaxPRR}
\begin{equation}
MI_{ij} = -\frac{1}{2}\log\left[1-\frac{\Sigma_{ij}\Sigma_{ji}}{\Sigma_{ii}\Sigma_{jj}}\right]
\end{equation}
This can be computed quickly and does not require us to find the maxent model itself. 

As we grow the network of constrained correlations, the algorithm used to compute a particular maximum entropy model takes more and more time to run. We can speed up the algorithm by taking advantage of the fact that it iterates over one edge at a time, allowing us to rewrite the algorithm using results for low rank matrix updates. Even with this speedup, adding edges to the graph takes longer and longer. As we create more opportunities to add loops, we have fewer opportunities to use the mutual information trick and have to run the full maxent algorithm more and more. 

To make progress, we developed a method for predicting the entropy drop that would result from adding any edge to the network without actually running the maxent algorithm (See section 2). Past a certain number of edges added, in this case 50, we switch from exactly finding the entropy drop from adding each candidate edge to predicting it. This allows us to then add the edge with the largest predicted entropy drop and run the maxent algorithm only once, to find the model corresponding to that new network of constraints. This model is then used in the process of predicting all of the entropy drops at the next step, and the process repeats. We can even increase the number of edges we add at a given time as the network of constraints becomes more dense. We add one edge per step until 200 edges added, then two per step until 400 edges, then five per step until 600 edges, then 10 per step until 800 edges, then 20 per step until 1000 edges, then 50 per step for the remainder of the process.

We use one final speedup. Maxent Algorithm 2,\cite{Uhler2017GaussianGraphicalModelsAlgebraic} which we discussed in the main text, runs quickly for sparse networks of edges but takes longer as the network becomes more dense. Algorithm 1\cite{Uhler2017GaussianGraphicalModelsAlgebraic} is slower for sparse networks but becomes faster for denser and denser networks. At some point for medium network density, in our case chosen to be 1500 edges added out of 4950, we switch from Algorithm 2 to Algorithm 1. With all of these steps, we are able to build minimax networks all the way from the independent model with no edges constrained to the full model with all of them constrained in a reasonable amount of time, on the order of hours.

\section{Predicting the entropy drop from adding an edge.} 
We imagine wiggling one element of the covariance and seeing how the model entropy changes, while enforcing the constraints imposed by the entropy maximization procedure. We can perform a series expansion of the entropy change in terms of the error in the corresponding element of the covariance matrix. Here we denote the model covariance as $K$, with model precision $J = K^{-1}$, and the data covariance as $\Sigma$. We fit all diagonal elements of the covariance from the start, so we can assume $i \neq j$
\begin{equation}
    \Delta S = \frac{dS}{dK_{ij}}(K_{ij} - \Sigma_{ij}) + \frac{1}{2}\frac{d^2S}{dK_{ij}^2}(K_{ij} - \Sigma_{ij})^2 + ...
\label{ent_expansion}
\end{equation}
The entropy of a Gaussian probability distribution has the following form:\cite{Cover&Thomas}
\begin{equation}
    S = \frac{N}{2}\ln{(2\pi e)} + \frac{1}{2}\ln{(|K|))}
\end{equation}
Starting with the first order term:
\begin{equation}
    \frac{dS}{dK_{ij}} = \frac{\partial S}{\partial K_{ij}} + \sum_{(kl)\notin E^{*}, (kl)\neq(ij)}\frac{\partial S}{\partial K_{kl}} \frac{dK_{kl}}{d K_{ij}}
\end{equation}
Where $E^*$ is the network of covariances that are being fit, including the diagonal (meaning the variances). The elements of $K$ that are being fit are not free parameters as they are constrained to match their experimental values, whereas the edges not in $E^*$ are able to change, which is why we sum over them above. Let's look at the partial derivatives:
\begin{equation}
    \frac{\partial S}{\partial K_{ij}} = \frac{1}{2|K|}\frac{\partial |K|}{\partial K_{ij}}
\end{equation}
To obtain the derivative on the right hand side, we use the following result for the partial derivative of the determinant:\cite{MagnusMatrixDiffEq}
\begin{equation}
    \frac{\partial |K|}{\partial K_{ij}} = |K|\Tr(J\frac{\partial K}{\partial K_{ij}})
\end{equation}
The matrix $\frac{\partial K}{\partial K_{ij}}$ is equal to one for both the $ij$ and $ji$ elements, as those elements are really the same parameter due to the enforced symmetry of the covariance matrix. We can denote the $mn$-th element of this as:
\begin{equation}
    \left( \frac{\partial K}{\partial K_{ij}}\right)_{mn} = \delta_{mi}\delta_{nj} + \delta_{mj}\delta_{ni}
    \label{delta equation}
\end{equation}
With this, we have:
\begin{equation}
    \Tr(J\frac{\partial K}{\partial K_{ij}}) = \sum_m \sum_k J_{mk} (\delta_{ki}\delta_{mj} + \delta_{kj}\delta_{mi})
    = J_{ji} + J_{ij} = 2J_{ij}
\end{equation}
Using this, we have:
\begin{equation}
    \frac{\partial S}{\partial K_{ij}} = \frac{1}{2|K|}\frac{\partial |K|}{\partial K_{ij}} = \frac{1}{2|K|}|K|2J_{ij} = J_{ij}
\end{equation}
Plugging this into our total derivative expression, while noting that if $(kl) \notin E^*$, then $k\neq l$ as we fit the variances from the start:
\begin{equation}
    \frac{dS}{dK_{ij}} =  J_{ij} + \sum_{(kl)\notin E^{*},(kl)\neq (ij)}J_{kl} \frac{dK_{kl}}{dK_{ij}}
\end{equation}
Entropy maximization requires that $J_{kl} = 0$ for $(kl) \notin E^*$, so we see that the sum on the right hand side vanishes (assuming the total derivative term is well behaved), and we are left with:
\begin{equation}
    \frac{dS}{dK_{ij}} =  J_{ij}
\end{equation}
We can now move on the the second derivative term in our expansion. We see now that 
\begin{equation}
    \frac{d^2S}{dK_{ij}^2} =  \frac{dJ_{ij}}{dK_{ij}}
\end{equation}
We will proceed by calculating $\frac{dK_{ij}}{dJ_{ij}}$ then inverting.
We can rewrite the above as follows:
\begin{equation}
    \frac{dK_{ij}}{dJ_{ij}} = \frac{\partial K_{ij}}{\partial J_{ij}} + \sum_{(kl)\in E^*} \frac{\partial K_{ij}}{\partial J_{kl}} \frac{dJ_{kl}}{dJ_{ij}}
\end{equation}
For the partial derivative of an inverse matrix with respect to an element of the original matrix, the following result is known for a general matrix M:\cite{MagnusMatrixDiffEq}
\begin{equation}
    \frac{\partial M^{-1}_{ij}}{\partial M_{kl}} = -\left(M^{-1} \frac{\partial M} {\partial M_{kl}} M^{-1} \right)_{ij}
\end{equation}
Using this, and setting $M = J$ and $M^{-1} = K$, we have:
\begin{equation}
    \frac{\partial K_{ij}}{\partial J_{kl}} = -\left(K \frac{\partial J}{\partial J_{kl}} K \right)_{ij}
\end{equation}
Using similar notation as in Eqn. \eqref{delta equation} we have, for $k \neq l$:
\begin{equation}
    \frac{\partial K_{ij}}{\partial J_{kl}} = - \sum_m K_{im} \sum_n  (\delta_{mk}\delta_{nl} + \delta_{ml}\delta_{nk}) K_{nj}  = -K_{ik}K_{lj} -K_{il}K_{kj}
    \label{partial C partial J}
\end{equation}
For $k=l$ we have:
\begin{equation}
    \frac{\partial K_{ij}}{\partial J_{kl}} = - \sum_m K_{im} \sum_n  \delta_{mk}\delta_{nk} K_{nj} = -K_{ik}K_{kj} 
\end{equation}
Now we just need to find $\frac{dJ_{kl}}{dJ_{ij}}$, then we will have the full expression for the second derivative term. To make progress, we use the following trick: For an edge $(kl) \in E^*$, and $(ij) \notin E^*$ we have that $K_{kl}$ cannot change, as it is being fit, and:
\begin{equation}
    \frac{dK_{kl}}{d J_{ij}} = 0 = \frac{\partial K_{kl}}{\partial J_{ij}} + \sum_{(mn) \in E^*} \frac{\partial K_{kl}}{\partial J_{mn}} \frac{d J_{mn}}{d J_{ij}}
\end{equation}
We can solve the equation above to find $\frac{d J_{mn}}{d J_{ij}}$. We do this by treating the above as a matrix equation $A\vec{x} = \vec{b}$, where:
\begin{equation}
    \vec{b}_{e_1} = -\frac{\partial K_{e_1}}{\partial J_{ij}}, \vec{x}_{e_2} = \frac{\partial J_{e_2}}{\partial J_{ij}}, A_{e_1 e_2} = \frac{\partial K_{e_1}}{\partial J_{e_2}}
\end{equation}
and $e_1$ and $e_2$ are edges in $E^*$. We can obtain all of the elements in $A$ and $b$ using \eqref{partial C partial J}. 
Assuming A is invertible, we have:
\begin{equation}
    \frac{dJ_{kl}}{dJ_{ij}} = -\sum_{(mn)\in E^*}  A^{-1}_{(kl),(mn)}\frac{\partial K_{mn}}{\partial J_{ij}}
\end{equation}
We can summarize
\begin{equation}
    \frac{dK_{ij}}{dJ_{ij}} = \frac{\partial K_{ij}}{\partial J_{ij}} - \sum_{(kl)\in E^*} \frac{\partial K_{ij}}{\partial J_{kl}}\sum_{(mn)\in E^*}  A^{-1}_{(kl),(mn)}\frac{\partial K_{mn}}{\partial J_{ij}}
\end{equation}
We now have all that we need. Eqn. \ref{ent_expansion} becomes:
\begin{equation}
    \Delta S \approx J_{ij}(K_{ij} - \Sigma_{ij}) + \frac{1}{2}\left\lbrack\frac{\partial K_{ij}}{\partial J_{ij}} - \sum_{(kl)\in E^*} \frac{\partial K_{ij}}{\partial J_{kl}}\sum_{(mn)\in E^*}  A^{-1}_{(kl),(mn)}\frac{\partial K_{mn}}{\partial J_{ij}}\right \rbrack^{-1} (K_{ij} - \Sigma_{ij})^2
\end{equation}
We are evaluating this for an edge $(ij)$ that has not yet been added to the network, so $J_{ij} = 0$, meaning that the first order term in the error vanishes. With the elements of the matrix $A$ and the partial derivatives given previously, everything in this expression now depends only on the elements of the current model covariance $K$ (that is, the one found at the prior step without the edge $(ij)$ constrained),  and the data covariance $\Sigma$. While the final result may not look pretty, it can be numerically calculated much faster than it would take to actually find the true maxent model with the same edge added. A comparison of predicted versus actual entropy drops can be found in Fig. S1 and we find good agreement.

\section{Predicting the information gained from adding a constraint}
\begin{figure}[h!]
\centering
\includegraphics[width = \textwidth]{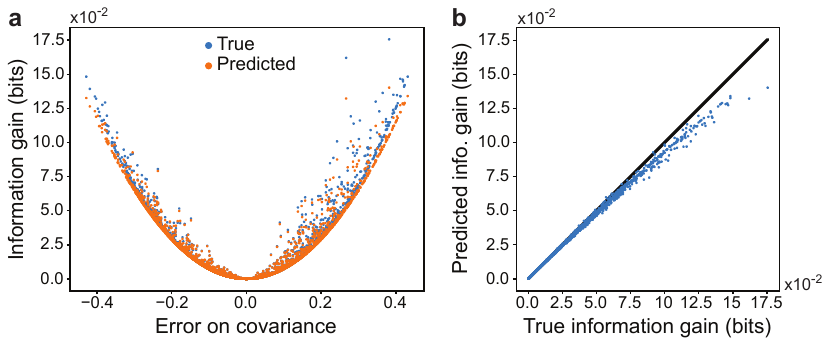} \\
\raggedright
\captionsetup{labelformat=empty}
{\spacing{1.25} \caption{\small \textbf{Fig. S\ref{fig:entropy_predictions} $|$ Information gained from adding a constraint can be predicted accurately.} \textbf{a}, Predicted and true information gains from adding each possible next constraint to the optimal network with 100 correlations fit, for the combined fMRI data, versus the error on the prediction of the corresponding covariance before it is fit. For each of true and predicted, each point represents one of the 4850 possible correlations that could be added to become the 101st edge in the network. \textbf{b}, Predicted information gain from adding each possible next constraint to the optimal network with 100 correlations fit, for the combined fMRI data, versus the true information gain. Each point represents one of the 4850 possible correlations that could be constrained. The diagonal line represents a slope of one. \label{fig:entropy_predictions}}}
\end{figure}
Here we compare our predictions for the true entropy drops, obtained by finding the corresponding maxent model and calculating its entropy, for each of the possible edges that could be added to the optimal network with 100 edges added for the combined fMRI data. In Fig. S\ref{fig:entropy_predictions}a we see the predicted and true information gains versus the error on the prediction of the corresponding element of the covariance before that edge was added. Notice the approximately quadratic shape, which is assuring given that the expansion in our prediction goes to second order in the error, with the linear term being zero. In Fig. S\ref{fig:entropy_predictions}b we see the predicted information gains plotted versus the true gains, finding good agreement.

\section{Comparison of our method of selecting edges to heuristics}
\begin{figure}[t!]
\centering
\includegraphics[width = 0.9
\textwidth]{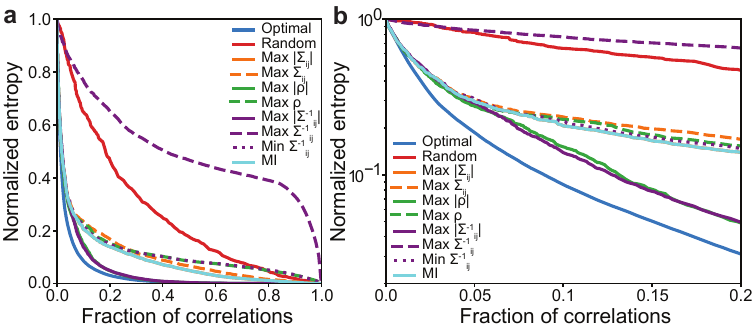} \\
\raggedright
\captionsetup{labelformat=empty}
{\spacing{1.25} \caption{\small \textbf{Fig. S\ref{fig:heuristics} $|$ Model entropy for heuristic methods.} \textbf{a}, Normalized model entropy of models of the combined fMRI data for various heuristics for selecting correlations to add to the model, compared to optimal and random methods. Here $\rho$ stands for partial correlation, $\Sigma$ is the data covariance, $\Sigma^{-1}$ is the data precision matrix, and MI stands for mutual information. \textbf{b}, Same as \textbf{a} but with log scale y-axis and for fraction of correlations between 0 and 0.2  
\label{fig:heuristics}}} 
\end{figure}

One could try to choose which correlations to add to the model as we build it using some heuristic, for example picking the largest correlations in magnitude, as we did in the main text. There are many other heuristics one could use. In figure S\ref{fig:heuristics} we compare the optimal, random, and max($|\Sigma_{ij}|$) methods from the main text to a list of other methods, all computed on the full data: max($\Sigma_{ij}$), max partial correlation in magnitude, max signed partial correlation, max($|\Sigma^{-1}_{ij}|$), max($\Sigma^{-1}_{ij}$), min($\Sigma^{-1}_{ij}$), and largest mutual informations, where $\Sigma^{-1}$ is the data precision matrix. For Gaussians, the partial correlation is $\frac{-\Sigma^{-1}_{ij}}{\sqrt{\Sigma^{-1}_{ii}\Sigma^{-1}_{jj}}}$, which measures the correlation between regions $i$ and $j$ when controlling for the indirect dependencies between them through all of the other regions in the brain. We see varying degrees of success, with the largest absolute partial correlations and the max($|\Sigma^{-1}_{ij}|$) working the best.

\section{Additional brain-system level structure analysis}

\begin{figure}[t!]
\centering
\includegraphics[width = 0.8
\textwidth]{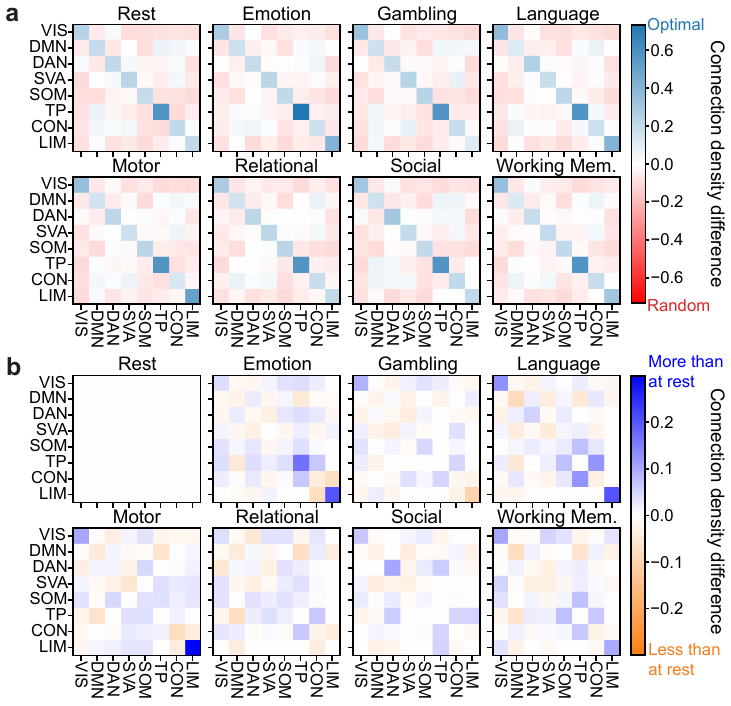} \\
\raggedright
\captionsetup{labelformat=empty}
{\spacing{1.25} \caption{\small \textbf{Fig. S\ref{fig:system_extras} $|$ System level structure of optimal models.} \textbf{a}, Difference in connection density between/within systems for the optimal model (blue) and random expectation (red) for tasks, with approximately 10\% of correlations \textbf{b}, Difference of connection density between each task and rest, with approximately 10\% of correlations. Blue corresponds to higher density in the task than at rest, and orange corresponds to less.
\label{fig:system_extras}}} 
\end{figure}

In main text Fig. 5a, we see the brain-system level structure of optimal models compared to the random expectation for all tasks. Specifically we examine the difference in connection density between/within systems for the optimal model and random expectation for networks with approximately 10\% of correlations. In Fig. S\ref{fig:system_extras}a we look at the same but for each task including rest. One consistent feature is that there are more edges within systems than would be expected at random for all systems, which indicates that the coarse-graining of regions into systems was reasonable. One might also be interested to see which regions are more or less likely to be informative for a certain task as compared to rest. In Fig. S\ref{fig:system_extras}b we see the difference of connection density between each task and rest, again for 10\% of correlations. We see some unique differences for each task, although we will not attempt to interpret them here.

\section{Model predictions relative to bootstrap errors}

\begin{figure}[h!]
\centering
\includegraphics[width = 0.9\textwidth]{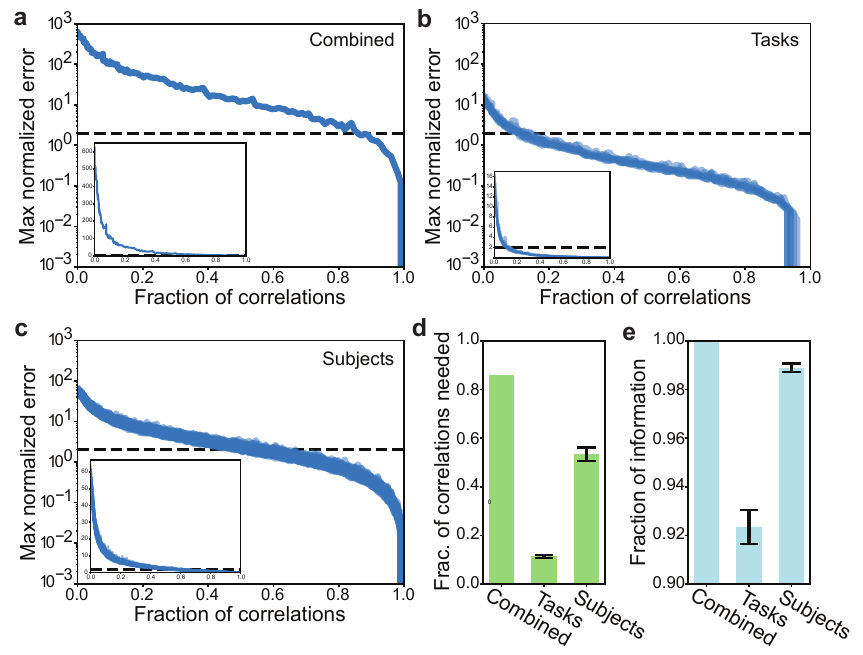} \\
\raggedright
\captionsetup{labelformat=empty}
{\spacing{1.25} \caption{\small \textbf{Fig. S\ref{fig:predict_boot} $|$ Maximum error on unconstrained covariances relative to bootstrap standard deviation.} \textbf{a}, Maximum of errors on unconstrained covariances divided by corresponding bootstrap standard deviation for data combined across both subjects and tasks. Dashed lines in a-c represent value of 2. Inset shows linear-linear plot of same results. \textbf{b}, Same as in (a) for data combined across subjects for a given task. \textbf{c}, Same as in (a) for data combined across tasks for a given subject. \textbf{d}, Fraction of edges added at which threshold of 2 is crossed. \textbf{e}, Fraction of possible entropy drop at the time when threshold is crossed. \label{fig:predict_boot}}}
\end{figure}

The high compressibilities seen in main text Fig. 3 suggest that with only a small fraction of correlations constrained, we should be able to accurately predict the rest. In order to get a sense for how accurate the predictions really are, we should measure the error relative to the inherent scale of uncertainties in the correlations themselves. To do this, we use the standard deviation of a correlation across bootstrap samples of the data (see Methods). We calculate the maximum error of our models for the combined data, for tasks, and for subjects on correlations not being fit in the models, divided by the bootstrap standard deviation of the corresponding correlation, with the results shown in Fig. S~\ref{fig:predict_boot}. We see that for the combined data, we need around 85.86\% of the edges to predict the remaining correlations to within two bootstrap standard deviations. This is perhaps not surprising given that each pair of regions may exhibit a significant correlation in at least one subject or task. If we focus on specific tasks, it takes a significantly smaller fraction, between 10.2\% and 12.73\% of the correlations added, to predict the others within errors. For individual subjects, it takes between 47.47\% and 60.61\% of the correlations. In all cases, we see a large decrease in the max error over bootstrap standard deviation for a relatively small fraction of edges, meaning that the sparse networks of covariances are able to give the majority of the reduction in prediction error. We would expect that once the predictions are within errors, the model is capturing most of the available information, and we see that this is indeed the case in Fig. S~\ref{fig:predict_boot}e. In summary, we find a large decrease in normalized prediction errors for a small fraction of correlations across subjects, tasks, and combined data, with the point at which the predictions are within errors changing significantly for different ways of slicing the data.

\section{Tolerances and Bootstrapping} To find reasonable error tolerances for the maxent algorithms above, we bootstrap sample the data 100 times and calculate the model covariance and precision matrix for each of those bootstrap sample timeseries. We then find the standard deviation of the elements of each matrix. We set the error tolerance as the smallest standard deviation divided by 100. For task models we used the smallest tolerances across tasks for all of the tasks due to a reasonably large difference in tolerances between tasks. For the subject models, we used each individual's tolerances.


\newpage

\section*{\large References}
\vspace{-30pt}
\noindent\rule{\textwidth}{.5pt}
\bibliography{refs}

\end{document}